\documentclass[floatfix,preprint,amsmath,amssymb,onecolumn,superscriptaddress,showpacs,longbibliography,11pt]{revtex4-2}
\usepackage{amsmath,amssymb}
\usepackage{graphicx}
\usepackage{appendix}

\usepackage[usenames,dvipsnames]{color}
\usepackage[colorlinks=true, citecolor=blue, linkcolor=blue, urlcolor=blue]{hyperref}
\usepackage{bm}
\usepackage{cleveref} 
\usepackage{braket}
\usepackage[nottoc,numbib]{tocbibind}
\usepackage{mathtools}
\usepackage{amsmath}
\usepackage{layouts}
\usepackage{caption}
\usepackage{subcaption}
\usepackage{graphicx}

\usepackage{lineno}
\usepackage{multirow}
\usepackage{dcolumn}

\newcommand{\pars}[1]{{(#1)}}

\setcitestyle{super}

\AtBeginDocument{}

\begin{document}

\title{Entanglement transition in deep neural quantum states}

\author{Giacomo Passetti}
\affiliation 
{Institut f\"ur Theorie der Statistischen Physik, RWTH Aachen University and JARA-Fundamentals of Future Information Technology, 52056 Aachen, Germany}

\author{Dante M.~Kennes}
\email{dante.kennes@mpsd.mpg.de}
\affiliation 
{Institut f\"ur Theorie der Statistischen Physik, RWTH Aachen University and JARA-Fundamentals of Future Information Technology, 52056 Aachen, Germany}
\affiliation 
{Max Planck Institute for the Structure and Dynamics of Matter, Center for Free-Electron Laser Science, Luruper Chaussee 149, 22761 Hamburg, Germany}

\date{\today}
\begin{abstract}
Despite the huge theoretical potential of neural quantum states, their use in describing generic, highly-correlated quantum many-body systems still often  poses practical difficulties. Customized network architectures are under active investigation to address these issues. For a guided search of suited network architectures a deepened understanding of the link between neural network properties and 
attributes of the physical system one is trying to describe, is imperative. Drawing inspiration from the field of machine learning, in this work we show how information propagation  in deep neural networks impacts the physical entanglement properties of deep neural quantum states. In fact, we link a previously identified information propagation phase transition of a neural network to a similar transition of entanglement in neural quantum states.
With this bridge we can identify optimal neural quantum state hyperparameter regimes for representing area as well as volume law entangled states. The former are easily accessed by alternative methods, such as tensor network representations, at least in  low physical dimensions, while the latter are challenging to describe generally due to their extensive quantum entanglement.  This advance of our understanding of network configurations for accurate quantum state representation helps to develop effective representations to deal with volume-law quantum states, and we apply these findings to describe the ground state (area law state) vs. the excited state (volume law state) properties of the prototypical next-nearest neighbor spin-1/2 Heisenberg model.
\end{abstract}
\maketitle


\section*{Introduction}
In recent years neural quantum states (NQS) have gained tremendous attention as a new promising technique for the exact representation of quantum many-body wave functions \cite{Carleo2017,Deng2017,Glasser2018,Kaubruegger2018,Choo2019,Fabiani2019,Schmitt2020,Fabiani2021,Astrakhantsev2021,Bukov_2021,Roth2022,Reh2023,sharir2022,Ceven_2022,Zou_2022,Szoldra_2023,Liu_FF_NN_2023,Humeniuk_2023, Nomura_2024}.
Initial studies focused on shallow architectures, such as Restricted Boltzmann Machines (RBM)\cite{Carleo2017, Deng2017}. Remarkably, these studies demonstrated how RBMs can efficiently represent quantum states with volume law entanglement in certain special cases\cite{Deng2017, Gao2017, Glasser2018, Levine_2019, sharir2022, Sun2022}. 
This was a significant advancement in the research of numerical ans\"{a}tze for representing complex quantum many-body states and their unique properties elevated NQSs as a potential complementary numerical technique to tensor network-related methods or others\cite{sharir2022}, which are often bound by low-entanglement constraints.

However, there exists also known limitations of shallow network architectures such as RBMs, which have been rigorously characterized \cite{Gao2017, Levine_2019}. In this context is was shown that RBMs cannot efficiently represent quantum states generated by generic polynomial-size quantum circuits. Therefore, it is nowadays well-established that shallow networks are insufficient to address generic problems of quantum many-body physics \cite{Gao2017, Levine_2019}. To remedy these limitations of shallow architectures using deep network architectures was established. Here, it has been demonstrated that increasing a network's depth, introducing more hidden layers, enables a higher efficiency of representation, for generic states in many cases\cite{Gao2017, Carleo2018, Levine_2019}. In fact, deep neural networks are known to exhibit an exponential increase in expressivity as the depth of the network increases\cite{Raghu2016}; a property that is crucial for their ability to model complex, high-dimensional functions\cite{Georgevici2019, Sarker2021, Chun_2023}. 
To train these deep networks efficiently a sophisticated network architecture, such as Convolutional Neural Networks (CNN) or Recurrent Neural Networks (RNN) can be beneficial\cite{Levine_2019}.

However, there is a tradeoff. (i) Deep neural networks are computationally expensive, require large amounts of data for training and are delicate to handle due to the large number of layers and (hyper-)parameters\cite{Chun_2023, Raza_2021}. (ii) Despite the high potential of the method, it has been recently pointed out that some complex states are hard to learn for NQS \cite{Lin2022, Passetti_2023}. (iii) Unlike traditional algorithms with rather transparent decision-making processes, the inner workings of deep neural networks are often considered 'black boxes', and  the characterization of how deep neural networks make decisions remains an open problem\cite{Brocki_2022}.

To remedy (iii) a deepened theoretical understanding of neural networks and their workings needs to be established and has gained intense attention under the general umbrella of ''understandable AI''. To address this problem a physics-inspired statistical field theory approach has been successfully applied\cite{Poole2016} within a mean field approximation. Within such a mean field analysis it was understood that tuning hyperparameters can trigger an information phase transition. This transitions separates an ordered from a disordered phase, in the sense that two inputs are either correlated or not by the network, in the ordered and disordered phase, respectively. In a physics-inspired language this corresponds to the emergence of a mean field order parameter. Deep in either the ordered or disordered phase the number of layers required to represent this correlation is exponentially small, while the number of involved layers diverges at the transition. This is akin to the behavior of the fluctuations at second order phase transitions in physical systems. This motivates to use networks tuned to this transition point in order to harness the full power of deep neural networks \cite{Poole2016, Raghu2016,Schoenholz2017}. 
Therefore, understanding these phase transitions and other properties of deep neural networks is key to comprehending their behavior, to demystify their inner processes and last but not least to improve their performance\cite{Poole2016, Schoenholz2017, Tamai_2023}.

As we aim to design state-of-the-art deep learning architectures for the description of quantum many-body physics, it becomes essential to adapt the existing knowledge from the broader machine learning community. More specifically, there is a pressing need to fine-tune neural networks to achieve efficient and advantageous representations of quantum many-body wave functions. In computational physics, representability of wavefunctions is often linked to their entanglement properties. For local, gapped Hamiltonians it has been shown that ground states exhibit low-entanglement restricted by the so called area law \cite{Eisert2010}; limiting the entanglement between two subparts of a system to the surface they share. Excited states generically feature a much more unfavorable entanglement scaling with system size described a volume law\cite{Bianchi2022}.

In this paper, we show that ordered to chaotic deep network phase transitions in the sense of Refs.~\onlinecite{Poole2016, Schoenholz2017, Tamai_2023} transduces a phase transition in the physical properties of corresponding deep NQSs.
In the chaotic vs. ordered phase of the neural network, the behavior of the entanglement entropy scaling of the corresponding NQS is markedly different in the sense that the ordered and the chaotic phase of the network can best represent area-law and volume-law states, respectively. Our work, thus, provides a reliable prescription on how to initialize a deep NQS with the desired scaling properties. We put our deepened understanding of the network's properties to use by demonstrating that the tuning of the network properties to either the ordered or chaotic phase allows us to better approximate the energy of physical ground states or excited states, respectively. We choose the physically relevant $J_{1}-J_{2}$ Heisenberg model and its ground state (area law) and mid-spectrum excited state (volume law) as our work horse to illustrate this point.

\section*{Main}
\subsection*{Ordered to Chaotic Phase Transition in Deep Neural Networks}
We briefly review some aspects of  the mean field formalism developed to study phase transitions in infinitely deep Feed Forward Neural Network (FFNN) as far as they are relevant to understanding our new findings. For further details, we refer to Refs.~\cite{Poole2016, Schoenholz2017}.
We consider a deep FFNN defined by a composition of $\mu \in \mathbb N$ layers where the $l$th network layer, $1 \le l \le \mu$, applies the transformation 
\begin{eqnarray}\label{eq::architecture}
\begin{split}
    z_{i}^\pars{l} = \sum_{j}W_{ij}^\pars{l}y_{j}^\pars{l} + b_{i}^\pars{l}, \qquad y_{i}^\pars{l+1} = \phi(z_{i}^\pars{l}), \qquad l = 1, \dots , \mu 
\end{split}
\end{eqnarray}
from input $y^\pars{l}_{i}$ to output signal $y^\pars{l+1}_{i}.$
This is an affine transformation with weight matrix (or kernel) $W^\pars{l}$, bias $b^\pars{l},$ and a nonlinear activation function $\phi.$
The networks weights and biases are drawn from zero mean Gaussian distribution such that 
$W^{l}_{ij}\sim N(0, \sigma_{w}^{2}/N_{l-1})$ and $b^{(l)}_{i} \sim N(0, \sigma_{b}^{2})$.

\begin{figure}
    \centering
    \begin{subfigure}[t]{0.49\textwidth}
        \centering
        \includegraphics[width=\linewidth]{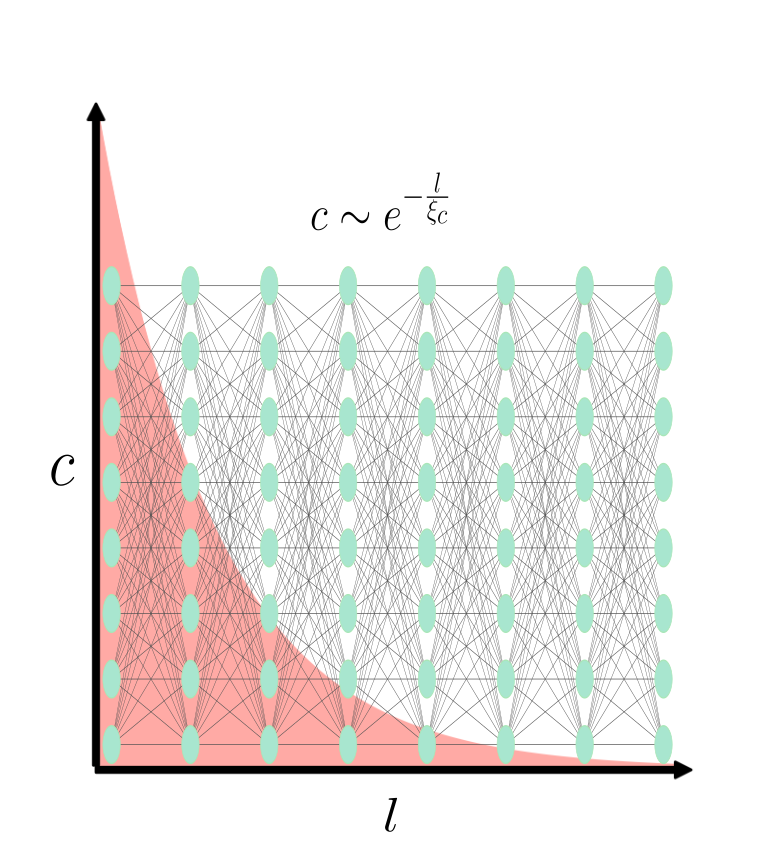} 
        \caption{} \label{fig:1a}
    \end{subfigure}
    \hfill
    \begin{subfigure}[t]{0.49\textwidth}
        \centering
        \includegraphics[width=\linewidth]{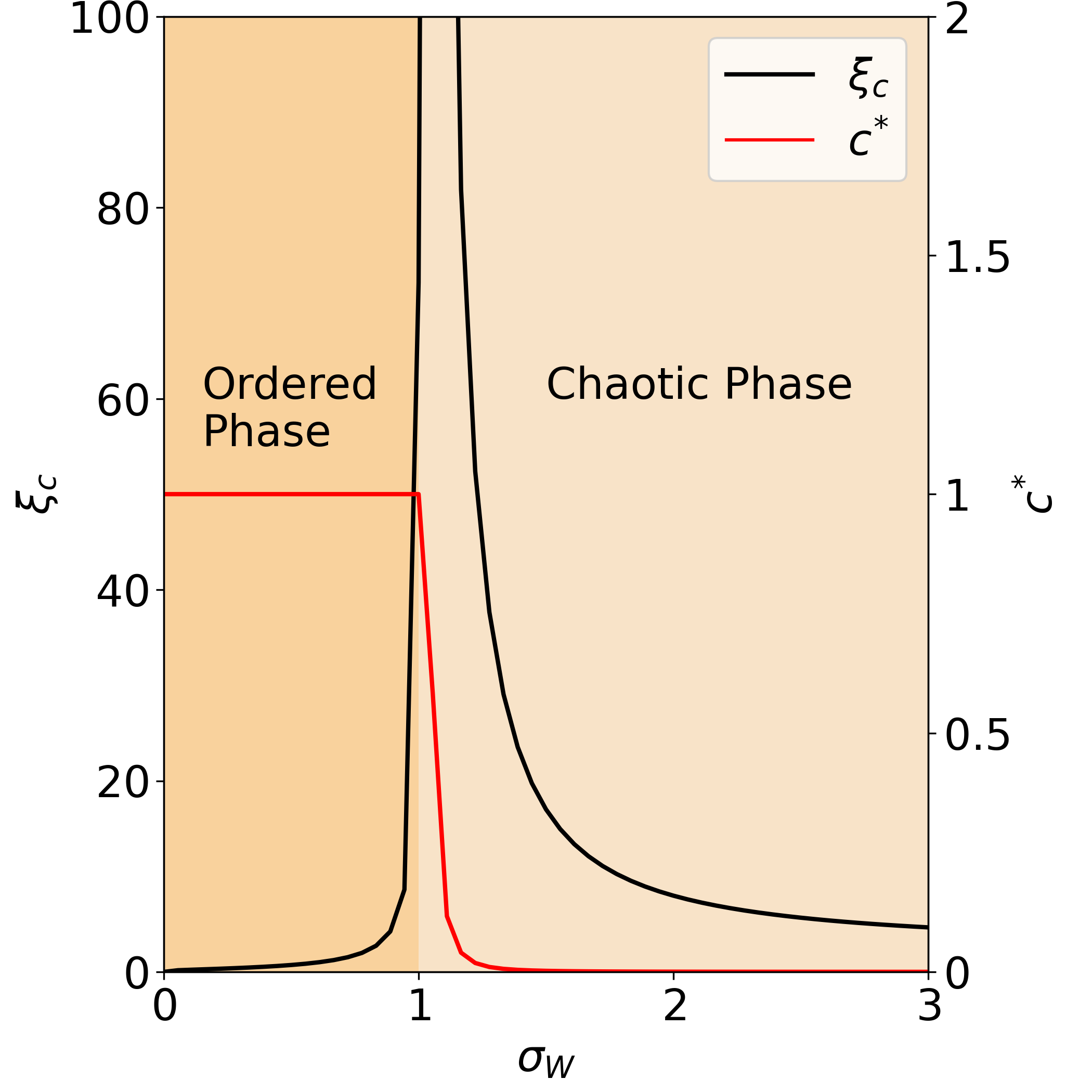} 
        \caption{} \label{fig:1b}
    \end{subfigure}
\caption{(a) Qualitative behavior for the propagation of the correlation function $c$ (defined by Eq.~\eqref{eq::network_correlation}) through a deep FFNN, the pink area  illustrates the typical exponential
decay characterized by the decay length $\xi_{c}$.
(b) Order parameter $c^{*}$ (red line) and 
Dependence of the decay length $\xi_{c}$ (black line) on $\sigma_{w}$, in the asymptotic limit of infinitely many layers. At the transition point from the ordered phase to the chaotic phase tracked by $c^{*}$, the decay length $\xi_{c}$ diverges. The figure shown is obtained by setting $\sigma_{b} = 0.01$.}\label{fig:1}
\end{figure}

Considering an initial input vector $z^{(0)}_{i ; \alpha}$ it has been analyzed how its magnitude is propagated along the network \cite{Poole2016}.
This, in fact, is mathematically equivalent to monitoring the second moment of the distribution, which at each layer $l$ reads 
$    \mathbb{E}\left[ z^{(l)}_{i; \alpha}z^{(l)}_{j; \alpha}\right] = q^{l}_{\alpha}\delta_{ij},
$
 where 
\begin{equation}\label{eq:recurrence_q}
    q^{l}_{\alpha} = \sigma_{w}^{2}\int \mathcal{D}z\phi^{2}\left(\sqrt{q_{\alpha}^{l-1}}z\right) + \sigma_{b}^{2}
\end{equation}
with  $ \int \mathcal{D}z= \frac{1}{\sqrt{2\pi}} \int dz e^{-\frac{1}{2}z^{2}}$.
For any arbitrary choice of $\sigma_{w}^{2}$ and $\sigma_{b}^{2}$ and bounded activation functions 
$\phi$ Eq.~\eqref{eq:recurrence_q} has a fixed point at $q^{*} = \lim_{l\rightarrow \infty} q_{\alpha}^{l}$ \cite{Poole2016}.
To quantify how two different input vectors get correlated across each layer of the network, one can consider two different initial input vectors $z^{(0)}_{i; \alpha}$ and $z^{(0)}_{i ; \beta}$, and look at the covariance of the layer defined pre-activation vectors $\mathbb{E}\left[z^{(l)}_{i ; \alpha}z^{(l)}_{j; \beta}\right] = q_{\alpha\beta}^l\delta_{ij}$.
The covariance function $q_{\alpha\beta}^{l}$ is then related to the correlation function $c^{l}_{\alpha\beta} = q^{l}_{\alpha\beta}/\sqrt{q^{l}_{\alpha}q^{l}_{\beta}}$
and can be evaluated through the recurrence relation
\begin{equation}
    q^{l}_{\alpha\beta} = \sigma_{w}^{2}\int \mathcal{D}z_{1}\mathcal{D}z_{2}\phi\left(u_{1}\right)\phi\left(u_{2}\right) + \sigma_{b}^{2}
\end{equation}
with $u_{1}=\sqrt{q_{\alpha}^{l-1}}z_{\alpha}$ and $u_{2}=\sqrt{q_{\beta}^{l-1}}\left(c_{\alpha\beta}^{l-1}z_{1}+\sqrt{1-(c_{\alpha\beta}^{l-1})^{2}}z_{2}\right)$.
It has been shown how the fix point $c^{*} = \lim_{l\rightarrow \infty }c^{l}_{\alpha\beta}$ allows to distinguish between an ordered phase characterized by $c^{*} = 1$ and a chaotic phase where \cite{Poole2016} $c^{*} = 0$.
As pointed out by Schoenholz et al.\cite{Schoenholz2017} correlations decay exponentially, see Fig.~\ref{fig:1} a), through a network to their asymptotic value according to 
\begin{equation}\label{eq::network_correlation}
c = |c^{l}_{\alpha\beta}- c^{*}| = e^{-l/\xi_{c}}.
\end{equation}
The behaviour of $\xi_{c}$ is reported in figure \ref{fig:1} b). We find, in correspondence to the information phase transition between ordered and chaotic phase,  a divergence of the correlation decay length\cite{Schoenholz2017}.

\subsection*{Deep feed forward Neural Quantum States}

We apply the general FFNN architecture presented in the previous paragraph to define a deep NQS describing spin-1/2 quantum states.
The weight matrices, introduced in Eq.\eqref{eq::architecture}, are now complex-valued, with elements still drawn from a Gaussian distribution. For simplicity, we restrict to the case of $\sigma_{b} = 0$.
The input vectors represent configurations of a spin chain, with a total number of spins $L$.
The last layer of the network is transformed into probability amplitudes using an exponential sum. Details of this implementation are presented in the Methods section.

\begin{figure}
    \centering
    \begin{subfigure}[t]{0.45\textwidth}
        \centering
        \includegraphics[width=\linewidth]{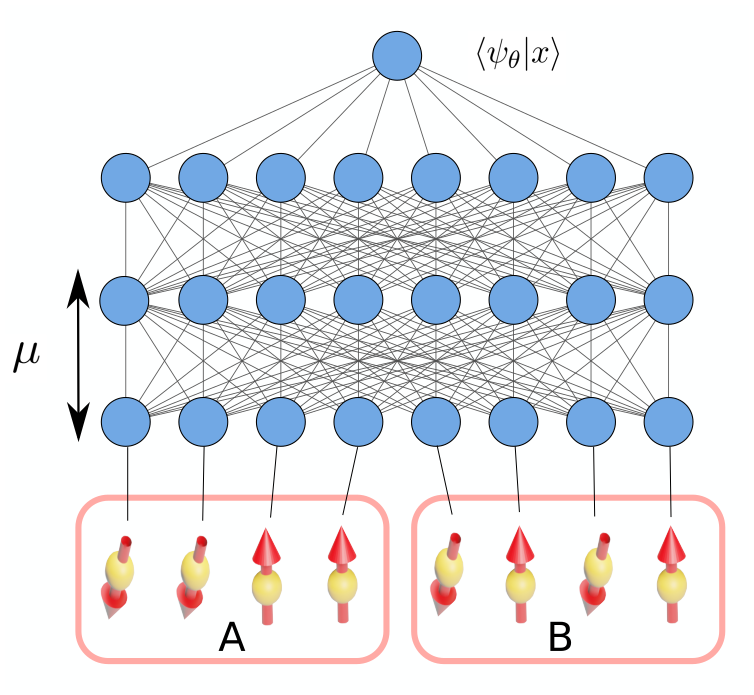} 
        \caption{} \label{fig:2a}
    \end{subfigure}
    \hfill
    \begin{subfigure}[t]{0.45\textwidth}
        \centering
        \includegraphics[width=\linewidth]{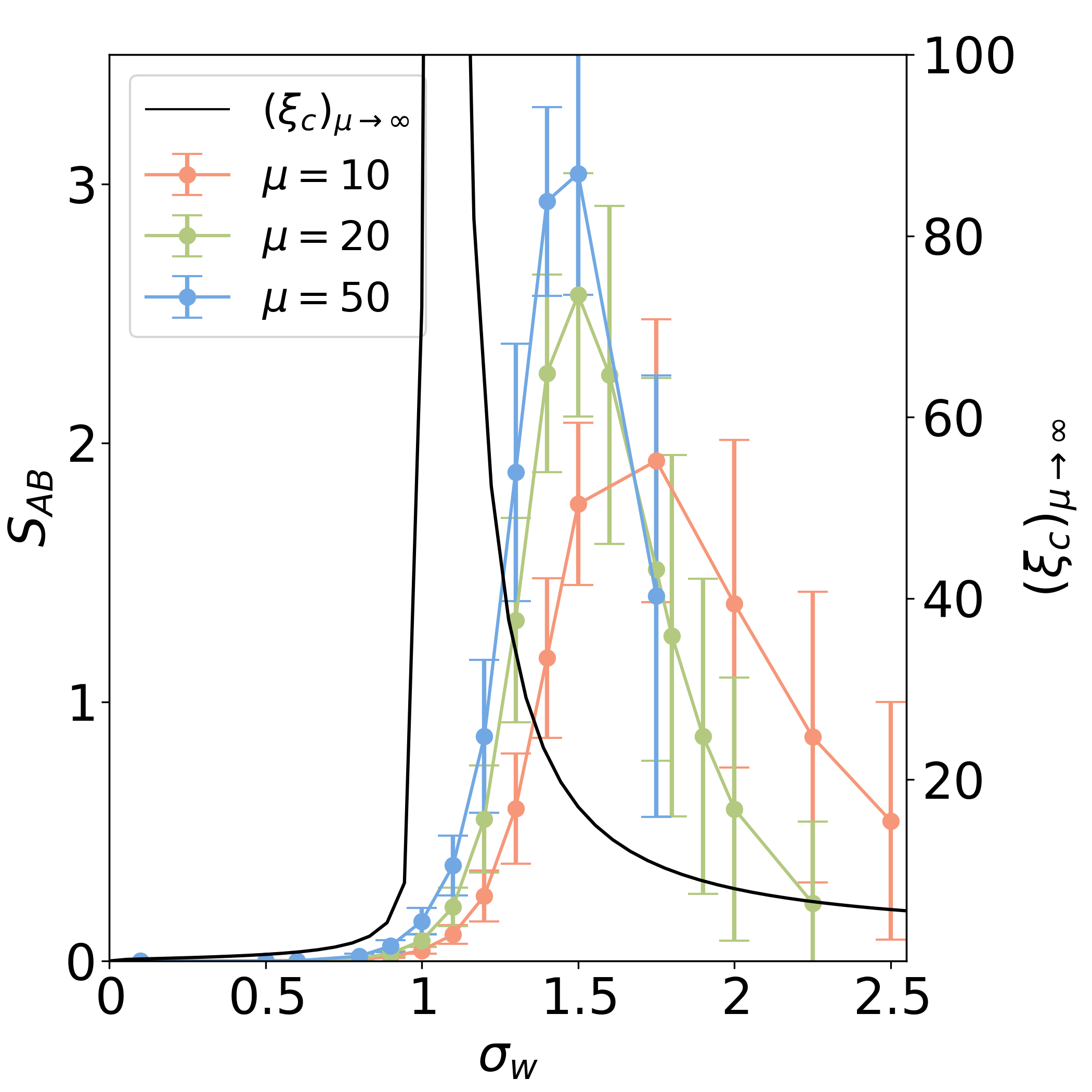} 
        \caption{} \label{fig:2b}
    \end{subfigure}
\caption{(a) Schematics of the setup considered, a deep FFNN represents the wave function for a spin chain. We consider the bipartition of the A and B 
halves of the chain, as sketched in the figure. 
(b) Bipartite entanglement entropy $S_{\mathrm{AB}}$ and correlation decay length $\xi_{c}$ varying $\sigma_{w}$, for different values of network layers $\mu$ (coloured lines, scale shown by left vertical axis).
We see that upon increasing the network depth, going toward the limit of an infinitely deep network, the profile of the entanglement entropy converges toward that of $\xi_{c}$ (black line, scale shown by the right vertical axis). 
Each data point has been averaged over $10^{3}$ independent realizations; the error bar corresponds to the standard deviation.}\label{fig:2}
\end{figure}

As depicted in Fig.~\ref{fig:2}(a) we then consider a bipartition of the spin chain, where the subpartition $A$ of the chain contains the first $\frac{L}{2}$ spins of the chain and the subpartition $B$ is simply the complementary subpartition.
Following the standard prescription, from any given quantum state represented by $\ket{\psi_{\theta}}$ the associated density matrix $\rho_{\theta} = \ket{\psi_{\theta}}\bra{\psi_{\theta}}$ allows to write the bipartite entanglement entropy as 

$    S_{\rm AB} = -\text{Tr}\left[\rho_{\rm A} \log \rho_{A}\right], \quad \rho_{\rm A} = \text{Tr}_{\rm B}\left[\rho\right]$
where $\text{Tr}_{\rm B}$ is the trace operation with respect to the degrees of freedom of the subpartition $\text{B}$.
We begin our analysis by showing that for a given fixed physical system size $L$ we observe that the profile for the average bipartite entanglement entropy $S_{\rm AB}$ for the NQS associated to random neural networks, has a dependence on $\sigma_{W}$ that exhibit a peak at the information phase transition point of the network, Fig.~\ref{fig:2}(b). The height of the entanglement entropy peak increases together with the network depth $\mu$ considered as expected for a physical phase transition in the entanglement properties. The resulting profile obtained simulating finite network realizations, approaches the theoretical prediction of $\xi_{c}$ for increasing numbers of layers. 

\subsection*{Connection between entanglement and information propagation transitions}

\begin{figure}
    \centering
    \begin{subfigure}[t]{0.49\textwidth}
        \centering
        \includegraphics[width=\linewidth]{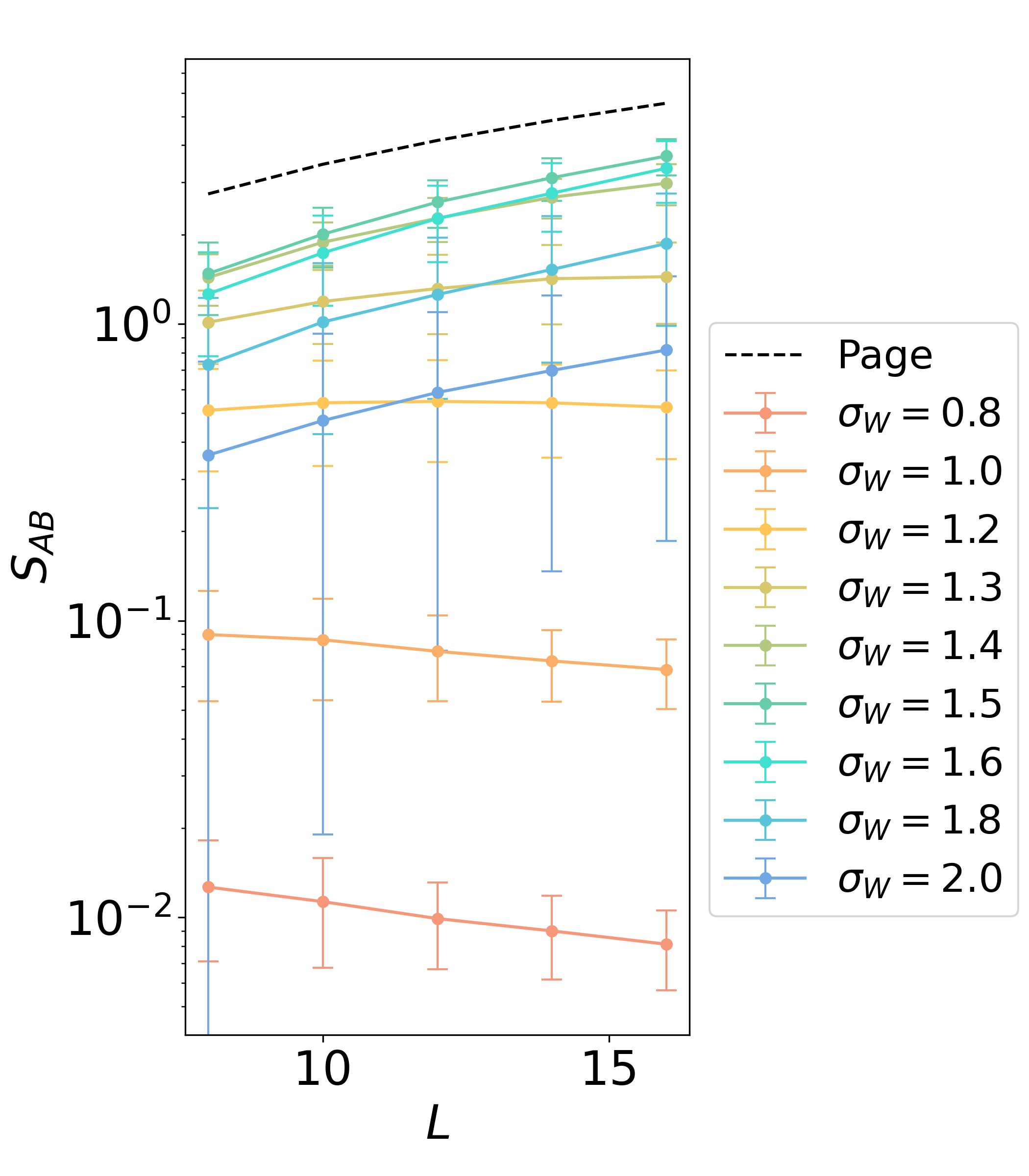} 
        \caption{} \label{fig:3a}
    \end{subfigure}
    \hfill
    \begin{subfigure}[t]{0.49\textwidth}
        \centering
        \includegraphics[width=\linewidth]{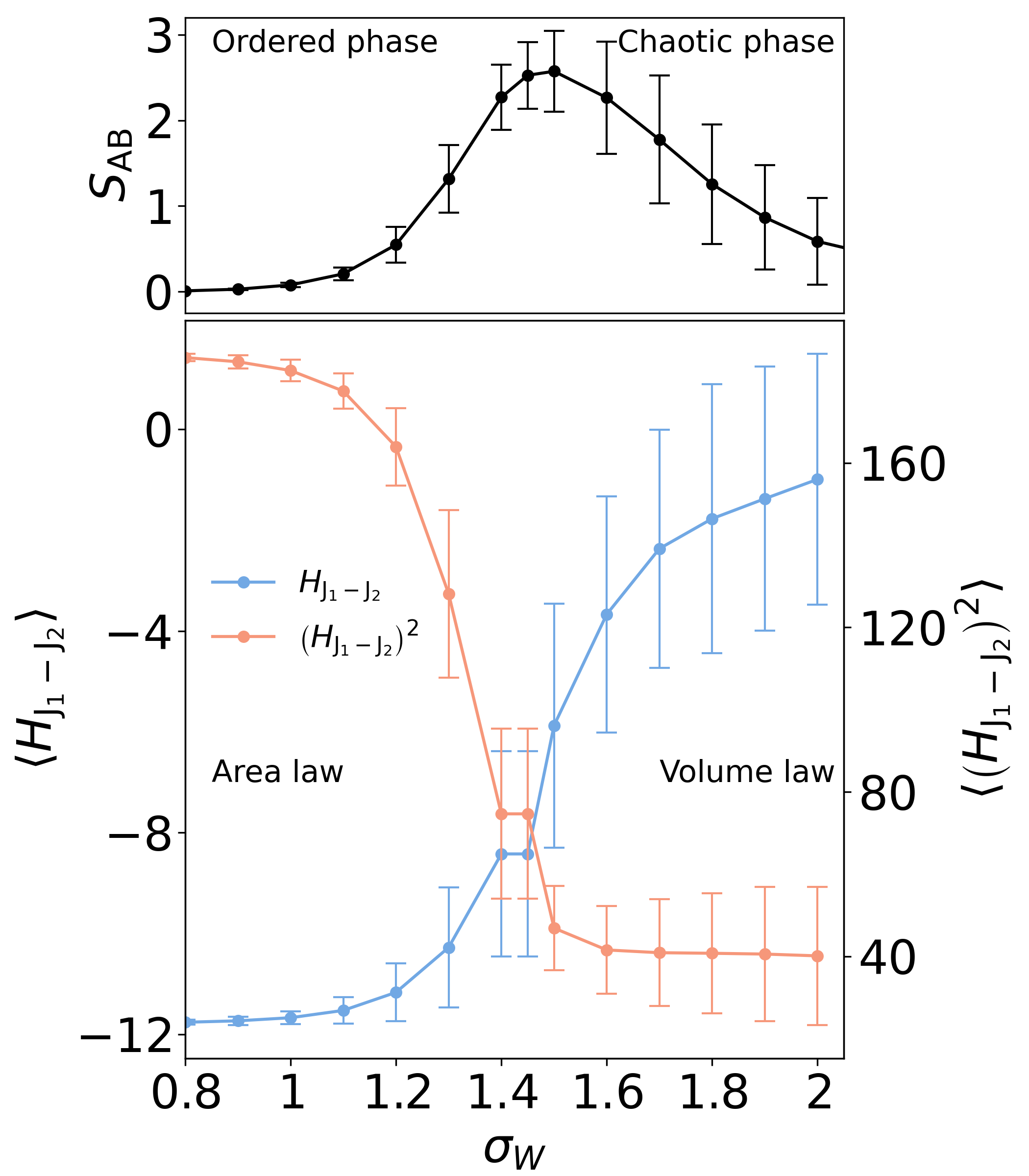} 
        \caption{} \label{fig:3b}
    \end{subfigure}
\caption{(a) Scaling of the bipartite entanglement entropy $S_{\mathrm{AB}}$ for deep FFNN of $\mu = 20$ layers.
Labelled different colors correspond to different value for $\sigma_{w}$, two behaviors are observed: For values of $\sigma_{W}$ in the ordered phase 
the bipartite entanglement entropy converges to a finite value, as for a 1D area law state. 
At $\sigma_{w}\approx 1.5$ the bipartite entanglement entropy almost saturates the Page value, and in the chaotic phase the entanglement displays still
a volume law scaling. 
(b) Top panel: Bipartite entanglement entropy $S_{\mathrm{AB}}$ for random networks with $\mu = 20$ layers, as function of $\sigma_{w}$. Bottom panel: Considering the same random network realizations we evaluate the energy expectation with respect to the hamiltonians $H_{J_{1}-J_{2}}$ (defined by Eq.~\ref{eq::hj1j2}) plotted with respect to the left axis, and $\left(H_{J_{1}-J_{2}}\right)^{2}$ (right axis).
We set $J_{1} = 1$ and $J_{2} = 0.2$.
Each data point has been averaged over $10^{3}$ independent realizations, the error bar corresponds to the standard deviation.}
\end{figure}\label{fig:3}

Next, we study how the bipartite entanglement entropy depends on the number of spins represented by $L$.
This scaling analysis allows to determine whether the quantum states that are represented by random network states satisfy a volume or an area law.
While in volume law states the $S_{\rm AB}$ grows linearly with the number of spins $L$, for an area law state $S_{\rm AB}$ should in general scale much weaker with systems size  (and approach a constant for example in the one-dimensional case considered later). 
The result of this analysis is presented in Fig.\ref{fig:3}.
In the same figure we compare the results with the Page entropy that corresponds to the entanglement entropy for a completely random quantum state and in the special case of bipartite entanglement is given by
\begin{equation}
    S_{\rm Page} = L\log (2) - \frac{1}{2}.
\end{equation}
As we can see, for values of the variance smaller than $\sigma_{w}\approx 1.4$, thus in the FFNN ordered phase, $S_{\rm AB}$ converges to a constant value, corresponding to an area law state.
Across the phase transition we see that in the chaotic case $S_{\rm AB}$ follows the same scaling behaviour as that of the Page entropy, thus following a volume law scaling.
In the chaotic phase, as the system approaches the phase transition point, $S_{\rm AB}$ increases and almost saturates to the Page value for $\sigma_{w}\approx 1.5$.
Deep FFNNs that are near the critical point, and thus can build up correlations between two distinct inputs, are then naturally mapped to deep random NQS which display volume law entanglement entropy.
To further emphasize how the deep network properties are directly translated into physical properties once the network is cast into the language of a  NQS we next consider the physical Hamiltonian operator
\begin{equation}\label{eq::hj1j2}
    H_{J_{1}-J_{2}} = J_{1}\sum_{\braket{ij}}\mathbf{S}_{i} \cdot \mathbf{S}_{j} + J_{2}\sum_{\braket{\braket{ij}}} \cdot \mathbf{S}_{i} \cdot \mathbf{S}_{j}
\end{equation}
with $\mathbf{S}_{i}$ the vector of spin-1/2 Pauli matrices.
This corresponds to the $J_{1}-J_{2}$ model, which is equivalent to the Heisenberg model with the addition of spin-spin interactions between next-to-nearest neighbour spins.
The ground state of this local Hamiltonian is known to be described by an area law for the entanglement entropy, up to logarithmic corrections when the system is tuned to be critical.
Highly excited states lying in the middle of the energy spectrum are instead typically described by a volume law for the entanglement entropy\cite{Bianchi2022}.
Because of this, the squared Hamiltonian operator $\left(H_{J_{1}-J_{2}}\right)^{2}$ is characterized by a volume law ground state.
Now we compare the $\sigma_{w}$ dependence of the energy expectation value for random deep NQS.
This is shown in Fig. \ref{fig:3b}, demonstrating a direct correspondence between the entanglement transition of the random NQS and the energy associated to the Hamiltonian as well as its square.
Importantly, the energy profile has an opposite behaviour in the two cases: (i) In the ordered phase, the network is closer in energy to the area law ground state of $H_{J_{1}-J_{2}}$.
(ii) In the chaotic phase, the network is closer in energy to the volume law ground state of $\left(H_{J_{1}-J_{2}}\right)^{2}$.
We demonstrated that the chaotic to order phase transition of the deep FFNN is mapped into a transition of the entanglement scaling properties for the associated NQS architecture, and that this allows to tailor the ansatz wave functions to be closer in energy to the physical quantum states that share the same entanglement properties.

\section*{Discussion}

We have investigated the entanglement characteristics of NQS constructed as deep FFNN with weight parameters drawn from a normal distribution. Our studies have revealed that the information propagation phase transition, originally explored in the context of random deep networks\cite{Poole2016, Schoenholz2017}, is directly related to the bipartite entanglement entropy as a function of the standard deviation of the network weights.
Our results provide a direct demonstration that the correlations established between independent input vectors and propagated through networks in proximity to the critical point are effectively translated into the generation of highly entangled quantum states. 

This discovery leads to a prescription for the preparation of initial wave function ans\"{a}tze that exhibit closer energy proximity to specific target ground state wave functions. In fact, we demonstrate that transitioning the network from an ordered phase to a chaotic phase facilitates the description of states with energy characteristics resembling area law ground states (in the ordered phase) or volume law states (in the chaotic phase).

Ground state search training protocols of challenging models are sensitive to the fine-tuning of  many hyper-parameters \cite{Passetti_2023}.
An understandable connection, such as the one we describe here,  is a central step forward in the NQS architecture development, where currently limits of learnability and representability are under heavy investigation \cite{Carleo2018, Passetti_2023}.
Our results pave the way for future investigations on potential boosts in the performance of optimization protocols that can efficiently target ground and excited states. 

\section{Methods}
We summarize here the prescriptions defining the NQS architecture studied in the main part.
We start from a conventional FFNN defined by 
\begin{eqnarray}\label{eq::architecture_nqs}
\begin{split}
    z_{i}^\pars{l} = \sum_{j}W_{ij}^\pars{l}y_{j}^\pars{l} + b_{i}^\pars{l}, \qquad y_{i}^\pars{l+1} = \phi(z_{i}^\pars{l}), \qquad l = 1, \dots , \mu.
\end{split}
\end{eqnarray}
For this, we consider the weight matrices appearing in Eq.~\eqref{eq::architecture} to be complex valued, such that
 $W_{ij}^{(l)} = w_{\mathrm{R}; ij}^{(l)} + iw_{\mathrm{I}; ij}^{(l)}$
is defined with $w_{\mathrm{R}; ij}^{(l)}$ and $w_{\mathrm{I}; ij}^{(l)}$ independently drawn from Gaussian distribution, setting $\sigma_{\rm R}^{2} = \sigma_{\rm I}^{2} = \sigma_{w}^{2}/N^{l-1}$. We consider configurations with no biases such that $b^{(l)}_{i} = 0$ and we choose the scaled exponential linear unit (SELU)\cite{Klambauer2017} as the activation function $\phi$.
The contraction of a network with a specific vector of the Hilbert space basis is then converted to the associated probability amplitude by applying an exponential sum, that, together with the definition in Eq.~\eqref{eq::architecture_nqs}, converts the last layer to the amplitude and is defined by
 $   \log \braket{x|\psi_{\theta}} = \log \sum_{i = 1}^{\alpha L}\exp\left[y^{(\mu)}_{i}\right], \quad  z^{(1)} = \sum_{j}W^{(1)}_{ij}x_{i}+b^{(1)}.$ 
The input vectors represent configurations of a spin chain, with a total number of spins $L$.
Therefore, each possible spin configuration consists of a vector $x \in \{0, 1\}^{L}$ , where $0$ and $1$ have been arbitrarily associated with spin down and up.
Since we restrict to system sizes $L$ that allow to numerically represent the base of the full Hilbert space, both the evaluations of the bipartite entanglement entropy reported in Figs.\ref{fig:2b}-\ref{fig:3a} and the energy expectation value reported in Fig.\ref{fig:3b} are obtained exactly.

\section*{Data availability}
All data supporting the findings of this study are available from the corresponding authors upon reasonable request.

\section*{Code availability}
The code supporting the findings of this study are available from the corresponding authors upon reasonable request.


\section*{ACKNOWLEDGEMENT} 
We acknowledge valuable discussions with Javed Lindner, Damian Hofmann, Claudia Merger and Lukas Grunwald.
We further acknowledge support from the
Max Planck-New York City Center for Non-Equilibrium
Quantum Phenomena. We acknowledge funding by the DFG under Germany's Excellence Strategy - Cluster of
Excellence Matter and Light for Quantum Computing (ML4Q) EXC 2004/1 - 390534769.  
NQS calculations have been performed using \textsc{NetKet}~3 \cite{NetKet3,NetKet2} with \textsc{jax} \cite{Jax2018}.
Computations were performed on the HPC system Ada at the Max Planck Computing and Data Facility (MPCDF), and with computing resources granted by RWTH Aachen University under project rwth0926.

\bibliography{bibliography.bib}

\section*{Author contributions}
GP performed the calculations and DMK supervised
the work. Both authors collaborated in writing the manuscript

\section*{Competing Interests}
The authors declare no competing financial or nonfinancial interests.

\end{document}